\documentclass[twocolumn,showpacs,superscriptaddress,preprintnumbers,amsmath,amssymb]{revtex4}

\usepackage{graphicx}
\usepackage{color}

\begin{document}

%=======================================================================================
\title{Unconventional plasmon-phonon coupling in graphene}

\author{Marinko Jablan}
\email{mjablan@phy.hr}
\affiliation{Department of Physics, University of Zagreb, Bijeni\v cka c. 32, 10000 Zagreb, Croatia}

\author{Marin Solja\v{c}i\'{c}}
\email{soljacic@mit.edu}
\affiliation{Department of Physics, Massachusetts Institute of Technology, 77 Massachusetts Avenue, Cambridge MA 02139, USA}

\author{Hrvoje Buljan}
\email{hbuljan@phy.hr}
\affiliation{Department of Physics, University of Zagreb, Bijeni\v cka c. 32, 10000 Zagreb, Croatia}

\date{\today}

\begin{abstract}
We predict the existence of coupled plasmon-phonon excitations in 
graphene by using the self-consistent linear response formalism. 
The unique electron-phonon interaction in graphene leads to unconventional 
mixing of plasmon and optical phonon polarizations. 
We find that longitudinal plasmons couple exclusively to transverse optical 
phonons, whereas graphene's transverse plasmons couple only to longitudinal 
optical phonons. This coupling can serve as a magnifier for exploring the electron-phonon 
interaction in graphene, and it offers novel electronical control over  
phonon frequencies. 
\end{abstract}

\maketitle
%\narrowtext
%\newpage

The interaction of electrons and crystal lattice vibrations (phonons) 
has fundamental implications on properties of materials and leads to diverse 
many-body phenomena such as superconductivity and charge-density waves. 
The electron-phonon interaction takes an unusual form in graphene, a recently 
discovered two-dimensional (2D) material \cite{Novoselov2004} (see, e.g., \cite{CastroNeto} 
for a review), and its implications are far from being explored. 
These include the breakdown of the Born-Oppenheimer approximation \cite{BO}, 
anomaly of the optical phonon \cite{Ando_anomaly}, and nonadiabatic Khon anomaly \cite{KhonAnomaly}. 
However, the interaction of collective electron excitations (plasmons) and 
optical phonons has not yet been presented for graphene. 
Plasmons in graphene are of fundamental scientific interest \cite{Wunsch2006,Hwang2007,
Mikhailov2007,Rana2008,Kramberger2008,Liu2010,Jablan2009}, but they also hold potential for 
technological applications (e.g., in the context of plasmon lasers \cite{Rana2008} and 
metamaterials \cite{Jablan2009}). 
Besides the ordinary longitudinal plasmons (transverse magnetic modes)
\cite{Wunsch2006,Hwang2007,Mikhailov2007,Rana2008,Jablan2009}, 
graphene also supports unusual transverse plasmons (transverse electric modes) 
\cite{Mikhailov2007}. 
The hybridization of plasmon and phonon modes is a striking manifestation of the 
breakdown of the Born-Oppenheimer approximation, because it 
occurs when phonons and electrons are on comparable energy scales. 
Plasmon-phonon coupling has been studied in bulk semiconductors \cite{Varga1965,Mooradian1966}, 
systems with reduced dimensionality (see e.g. \cite{Matz1981,Wu1985,Wendler1987,
Jalabert1989}), and in the context of graphene, plasmons were shown to couple to 
surface optical phonons of the substrate (e.g., SiC, which is a polar material) 
\cite{Hwang2010,Liu2010,Koch2010}. However, to the best of our knowledge, 
the phenomenon of plasmon-phonon coupling was not yet presented for an 
isolated 2D material. 
Here we predict the coupling of plasmons with intrinsic optical phonons
in graphene by using the self-consistent linear response formalism. 
We find that, in contrast to all other known systems in nature, longitudinal plasmons (LP) 
couple only to transverse optical (TO) phonons \cite{SarmaComment}, while transverse 
plasmons (TP) couple only to longitudinal optical (LO) phonons. 
The LP-TO coupling is stronger for larger concentration of carriers, 
in contrast to the TP-LO coupling (which is fairly weak). 
The former could be measured via current experimental techniques. 
Thus, plasmon-phonon resonance could serve as a magnifier for exploring 
the electron-phonon interaction, and for novel electronic 
control (by externally applied voltage) over crystal lattice 
vibrations in graphene.

\begin{figure}
\centerline{
\mbox{\includegraphics[width=0.50\textwidth]{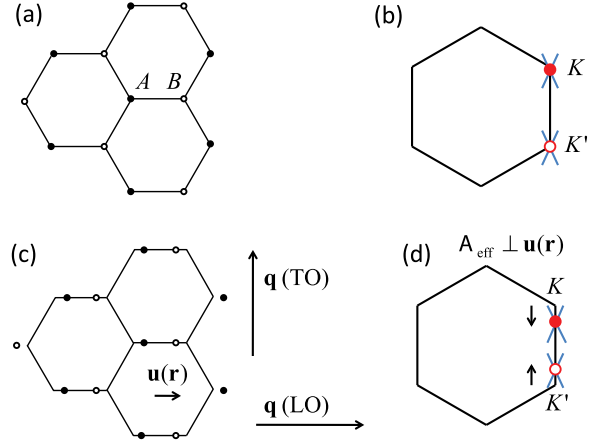}}
}
\caption{
(a) Schematic illustration of the lattice structure with two sublattices (A and B).
(b) The two degenerate Dirac cones are centered at K and K' points at the edge 
of the Brillouin zone. 
(c) A displacement of lattice atoms ${\bf{u}}({\bf r})$ is parallel (perpendicular) 
to the propagation wave vector ${\bf q}$ of a LO (TO) phonon.
(d) The displacement ${\bf{u}}({\bf r})$ creates an effective vector potential 
${\bf A}_{\textrm{eff}}$ {\em perpendicular} to ${\bf{u}}({\bf r})$ 
(the sign of ${\bf A}_{\textrm{eff}}$ for the K' point is opposite to 
that for the K point). 
}
\label{slika}
\end{figure}

The low-energy band structure of graphene consists of two degenerate Dirac cones 
at K and K' points of the Brillouin zone \cite{Wallace1954,Semenoff1984} 
[see Fig. \ref{slika}(a) and (b)], 
and the electron Hamiltonian around K point can be written as
\begin{equation}
H_{e} = \hbar v_F \mbox{\boldmath $\sigma$} \cdot {\bf k}
\label{Dirac} 
\end{equation}
where $v_F=10^6$~m/s, ${\bf k} = (k_x,k_y)=-i\mbox{\boldmath $\nabla$}$ is the 
wave-vector operator, 
$\mbox{\boldmath $\sigma$}=(\sigma_x,\sigma_y)$, and $\sigma_{x,y}$ are the Pauli 
spin matrices. We label the eigenstates of Hamiltonian $H_e$ by $| s, {\bf k} \rangle $ and 
the appropriate eigenvalues by $E_{s,\bf k}=s \hbar v_F |\bf k|$, where $s=1$ for 
the conduction band and $s=-1$ for the valence band. 
A technologically interesting property of graphene is that the concentration of electrons $n$,
and hence the Fermi level $E_F=\hbar v_F \sqrt{\pi n}$, can be changed 
via gate voltage \cite{Novoselov2004}.

The long-wavelength in-plane optical phonon branch in graphene consists of two modes (LO and TO) 
which are effectively dispersionless and degenerate at energy $\hbar\omega_0=0.196eV$ \cite{Ando_opt_ph, Ando_el_ph}. 
Let ${\bf{u}}({\bf R})=[{\bf u}_A({\bf R})-{\bf u}_B({\bf R})]/\sqrt{2}$ denote the 
relative displacements of the sub-lattice atoms $A$ and $B$ of a unit cell specified 
by a coordinate ${\bf R}$ [see Fig. \ref{slika}(c)]. 
Then, in the long-wavelength limit ${\bf R}$ can be replaced 
by a continuous coordinate {\bf r} and we have
\begin{equation}
{\bf{u}}({\bf r})=\sum_{\mu {\bf q}} \frac{1}{\sqrt{NM}}Q_{\mu {\bf q}}{\bf e}_{\mu {\bf q}}e^{i{\bf q}{\bf r}},
\label{ph-displace}
\end{equation}
where $N$ is the number of unit cells, $M$ is the carbon atom mass, 
${\bf q}=q(\cos\phi_{\bf q},\sin\phi_{\bf q})$ is the phonon wave vector, 
$\mu=L,T$ stands for the polarization, and the polarization unit vectors are 
${\bf e}_{L{\bf q}}=i(\cos\phi_{\bf q},\sin\phi_{\bf q})$, and 
${\bf e}_{T{\bf q}}=i(-\sin\phi_{\bf q},\cos\phi_{\bf q})$. 
The displacement vector ${\bf{u}}({\bf r})$ is parallel (perpendicular) to the 
phonon propagation wave vector ${\bf q}$ for LO (TO, respectively) phonons 
[see Fig. \ref{slika}(c)]. The phonon Hamiltonian is given by
\begin{equation}
H_{ph}=\frac{1}{2} \sum_{\mu {\bf q}} (P_{\mu \bf q}^{\dag} P_{\mu \bf q}+
\omega_0^2 Q_{\mu \bf q}^{\dag} Q_{\mu \bf q}),
\label{ph-Hamilt}
\end{equation}
where $Q_{\mu \bf q}$ and $P_{\mu \bf q}$ denote phonon coordinate and momentum. 
The electron-phonon interaction takes a peculiar form in graphene \cite{Ando_el_ph}
\begin{equation}
H_{e-ph} = -\sqrt{2} \frac{\beta \hbar v_F}{b^2} \mbox{\boldmath $\sigma$} \times {\bf{u(r)}},
\label{e-ph1}
\end{equation}
where $\mbox{\boldmath $\sigma$} \times {\bf{u}}=\sigma_x u_y-\sigma_y u_x$, $b=0.142$~nm is 
the nearest carbon atoms distance, and $\beta=2$. 
We find it convenient to write Eq. (\ref{e-ph1}) as
\begin{equation}
H_{e-ph} = L^2 F \sum_{\mu {\bf q}} {\bf j}_{\bf q}^{\dag} \times {\bf e}_{\mu {\bf q}} Q_{\mu {\bf q}}
\label{e-ph2}
\end{equation}
where ${\bf j}_{\bf q}=-e v_F L^{-2} \mbox{\boldmath $\sigma$} e^{-i{\bf q}{\bf r}}$ is the 
single-particle current-density 
operator, $L^2$ is the area of the system, $e$ is charge of the electron, and 
$F=\frac{\sqrt{2} \beta \hbar}{e b^2 \sqrt{NM}}$.

The electromagnetic field in the plane of graphene is completely described by 
the vector potential ${\bf A}=\sum_{\mu {\bf q}} {\bf e}_{\mu {\bf q}} A_{\mu {\bf q}} e^{i{\bf q}{\bf r}}$ 
(scalar potential is gauged to zero, time dependence is implicitly assumed, and 
$\mu=L,T$ denote polarizations). 
The interaction with Dirac electrons is obtained by substitution 
$\hbar {\bf k}\rightarrow \hbar {\bf k}+ e {\bf A}$ in Eq. (\ref{Dirac}),
which leads to 
\begin{equation}
H_{e-em}=e v_F \mbox{\boldmath $\sigma$} \cdot {\bf A} =
-L^2 \sum_{\mu {\bf q}} {\bf j}_{\bf q}^{\dag} \cdot {\bf e}_{\mu {\bf q}} A_{\mu {\bf q}}.
\label{e-em}
\end{equation}
By comparing Eqs. (\ref{e-ph1}) and (\ref{e-em}) 
it follows that electron-phonon interaction can be regarded 
as a presence of an effective vector potential 
\begin{equation}
{\bf A}_{\textrm{eff}}=F \sum_{\bf q} 
({\bf e}_{T {\bf q}} Q_{L {\bf q}} -
{\bf e}_{L {\bf q}} Q_{T {\bf q}})
e^{i{\bf q}{\bf r}},
\label{Aeff}
\end{equation}
that is, $H_{e-ph}=e v_F \mbox{\boldmath $\sigma$} \cdot {\bf A}_{\textrm{eff}}$. 
It is evident that ${\bf A}_{\textrm{eff}} \cdot {\bf{u}}({\bf r})=0$ that is 
the effective vector potential ${\bf A}_{\textrm{eff}}$ is perpendicular to 
${\bf{u}}({\bf r})$ as illustrated in Figs. \ref{slika}(c) and (d) (see also 
Ref. \cite{CastroNeto}), which is responsible for the mixing of polarizations in 
plasmon-phonon coupling. 

As a first pass, let us ignore the phonons and focus on the Hamiltonian $H=H_e+H_{e-em}$.
Without an external perturbation, the electrons in graphene fill the Fermi sea according 
to the Fermi distribution function $f_{s {\bf k}}$. 
A field $A_{\mu {\bf q}} (\omega)$ oscillating at frequency $\omega$ 
will induce an average current density (up to a linear order in the vector 
potential) 
\begin{equation}
\langle { J}_{\mu}({\bf q},\omega) \rangle = - \chi_{\mu}({\bf q},\omega)  A_{\mu {\bf q}} (\omega),
\label{current}
\end{equation} 
where the current-current response function (including 2-spin and 2-valley degeneracy) 
is given by \cite{PinesBook}
\begin{align}
\chi_{\mu}({\bf q},\omega)=4 L^2 \sum_{s_1s_2{\bf k}} & \frac{f_{s_1 {\bf k}}-f_{s_2 {\bf k}+{\bf q}}}{\hbar\omega+\hbar\omega_{s_1 {\bf k}}-\hbar\omega_{s_2 {\bf k}+{\bf q}}+i\eta} 
\nonumber \\
& \times |\langle s_1{\bf k}|{\bf j}_{\bf q} \cdot {\bf e}_{\mu {\bf q}}^*|s_2 {\bf k}+{\bf q}\rangle|^2.
\label{chi}
\end{align}
For the response function  $\chi_{\mu}({\bf q},\omega)$ we utilize the 
analytical expression from Ref. \cite{Principi2009}. 
The subtlety involved with the divergence in Eq. (\ref{chi}) is 
solved by subtracting from $\chi_{L}({\bf q},\omega)$ 
[$\chi_{T}({\bf q},\omega)$] the value $\chi_{L}({\bf q},\omega=0)$ 
[$\chi_{T}({\bf q} \to 0,\omega=0)$]
to take into account that there is no current response 
to the longitudinal [transverse] time [time and space] independent vector potential, 
see \cite{Principi2009, Falkovsky2007} for details. 
We would like to note that when working with the current-current response 
function, rather than with the density-density response function, the 
nature of the plasmon-phonon interaction (especially the mixing of 
polarizations as shown below) is far more transparent.

\begin{figure}
\centerline{
\mbox{\includegraphics[width=0.25\textwidth]{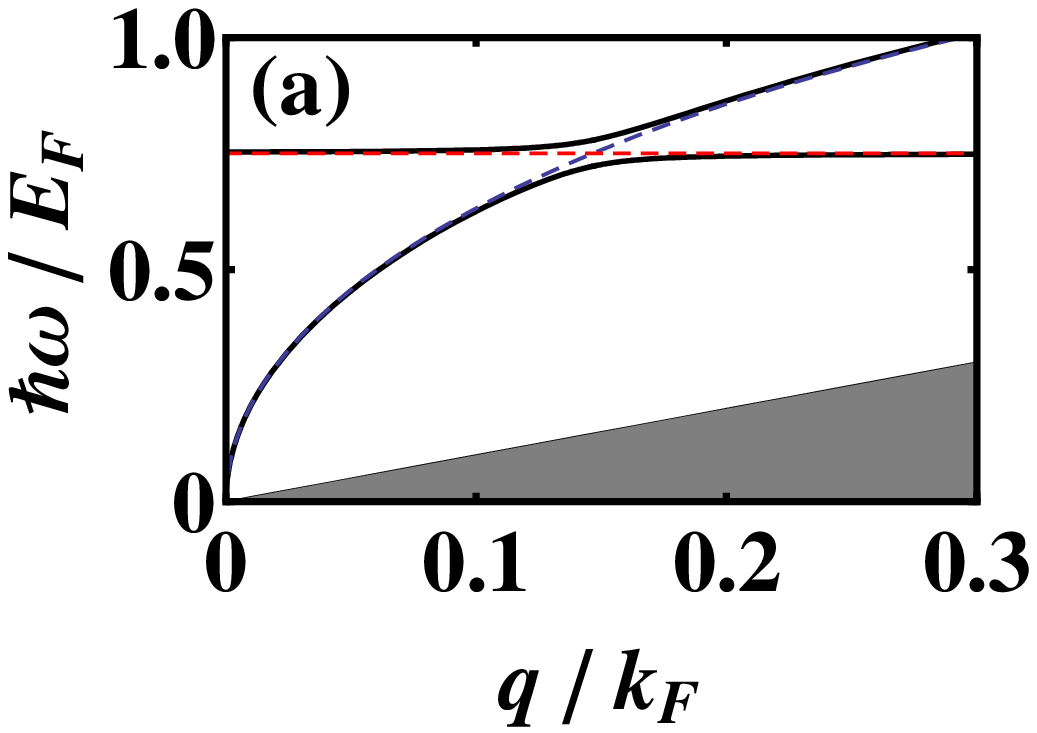}}
\mbox{\includegraphics[width=0.25\textwidth]{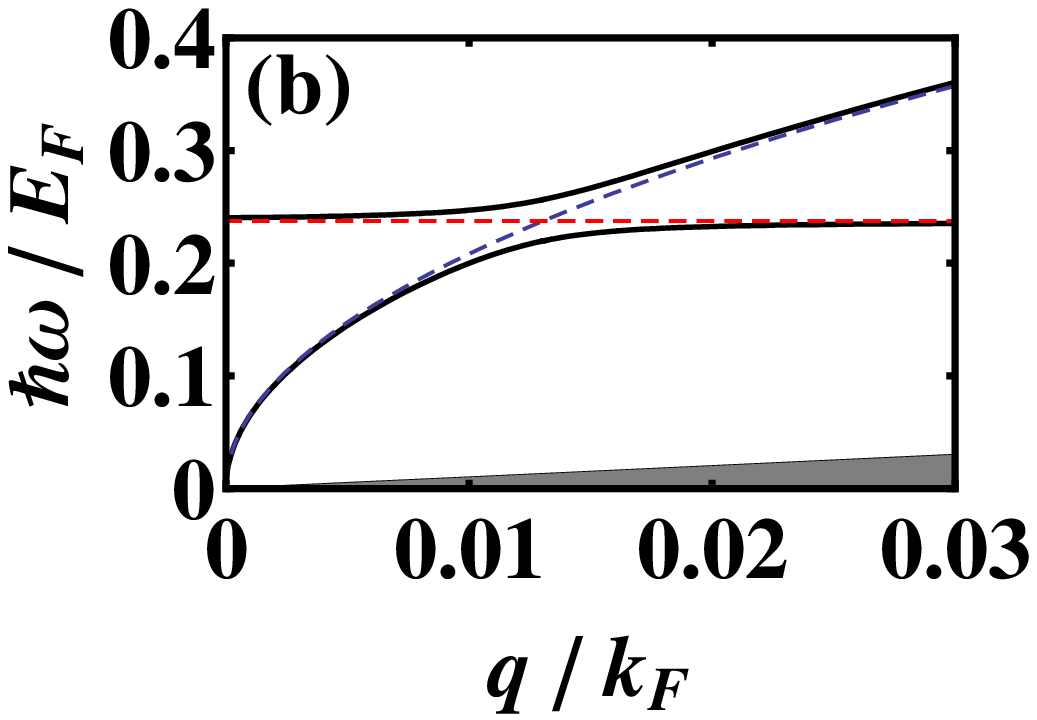}}
}
\caption{
Dispersion lines of hybrid LP-TO plasmon-phonon modes (solid lines) 
and of the uncoupled modes (dashed lines) for two values of doping:
(a) $n=5\times 10^{12}$~cm$^{-2}$, and 
(b) $n=5\times 10^{13}$~cm$^{-2}$. 
The hybridization is stronger for larger doping values. 
Grey areas denote the region of single-particle damping.
}
\label{fig1}
\end{figure}

\begin{figure}
\centerline{
\mbox{\includegraphics[width=0.25\textwidth]{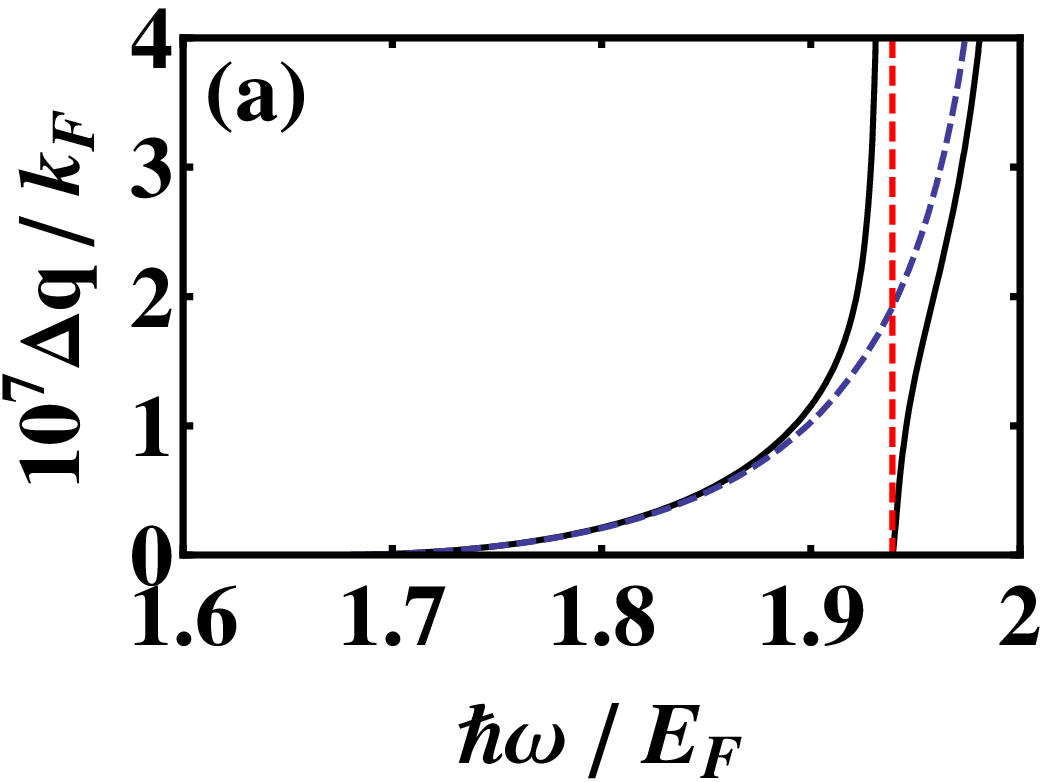}}
\mbox{\includegraphics[width=0.25\textwidth]{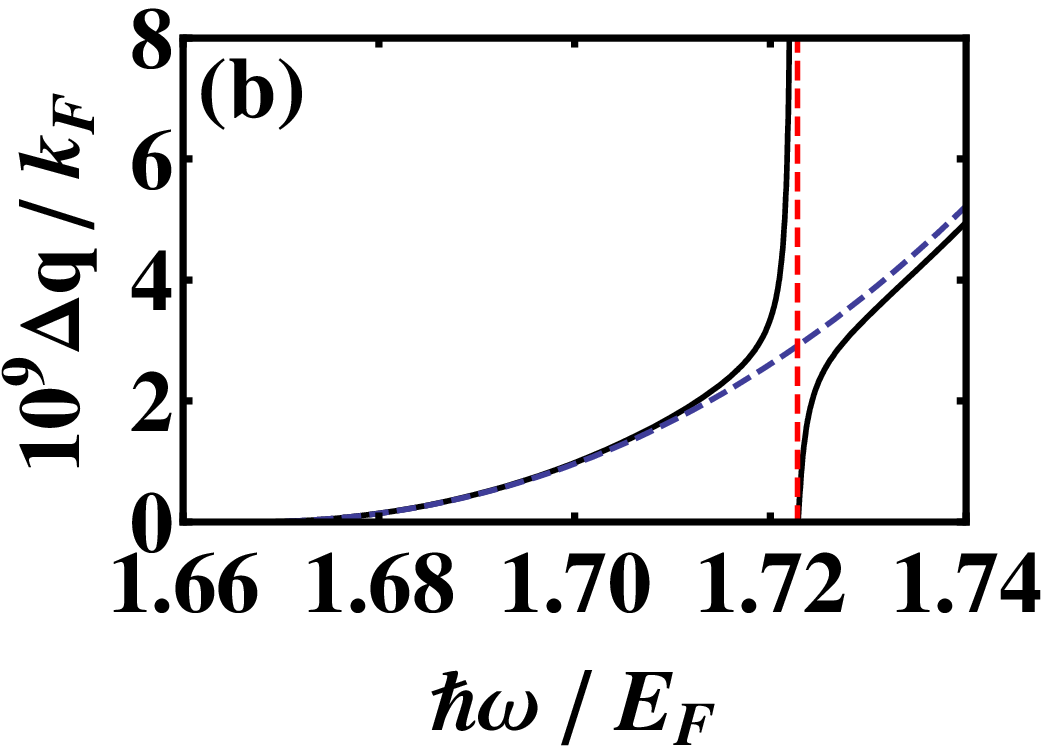}}
}
\caption{
Dispersion lines of hybrid TP-LO plasmon-phonon modes (solid lines) 
and of the uncoupled modes (dashed lines) for two values of doping:
(a) $n=7.5\times 10^{11}$~cm$^{-2}$, and 
(b) $n=9.5\times 10^{11}$~cm$^{-2}$. The plasmon-like dispersion 
is very close to the light line $q=\omega/c$; therefore, the ordinate shows 
$\Delta q=q-\omega/c$. 
}
\label{fig2}
\end{figure}

Next, it is straightforward to show from the Maxwell equations that an electric current 
oscillating in a two-dimensional plane will induce a vector potential
\begin{equation}
\langle A_{L {\bf q}}(\omega) \rangle = \langle { J}_L({\bf q},\omega) \rangle 
\frac{\sqrt{q^2-\omega^2/c^2}}{-2\omega^2\epsilon_0},
\label{Maxwell-long}
\end{equation} 
and
\begin{equation}
\langle A_{T {\bf q}}(\omega) \rangle = \langle { J}_T({\bf q},\omega) \rangle 
\frac{\mu_0}{2\sqrt{q^2-\omega^2/c^2}},
\label{Maxwell-trans}
\end{equation} 
where we have assumed that graphene is suspended in air and 
that there are no other sources present in space.  
This induced vector potential in turn acts on electrons in graphene through the interaction 
Hamiltonian $H_{e-em}$ which can result in plasmons - self-sustained collective oscillations of 
electrons. From Eqs. (\ref{current}) and (\ref{Maxwell-long}) we get the dispersion relation 
for longitudinal plasmons \cite{Wunsch2006,Hwang2007,Jablan2009}
\begin{equation}
1-\frac{\sqrt{q^2-\omega^2/c^2}}{2\omega^2\epsilon_0}\chi_L({\bf q},\omega)=0.
\label{plasmon-long}
\end{equation} 
From Eqs. (\ref{current}) and (\ref{Maxwell-trans}) we get the dispersion relation 
for transverse plasmons \cite{Mikhailov2007}
\begin{equation}
1+\frac{\mu_0}{2\sqrt{q^2-\omega^2/c^2}}\chi_T({\bf q},\omega)=0.
\label{plasmon-trans}
\end{equation} 
Longitudinal plasmons are also referred to as transverse magnetic modes since they 
are accompanied by a longitudinal electric ($E$) and a transverse magnetic field 
($B$) in the plane of graphene. Likewise transverse plasmons or transverse electric 
modes are accompanied by a transverse electric and a longitudinal magnetic 
field \cite{Mikhailov2007}. 
Dispersion relation of LP (TP) modes is shown by the blue dashed line in 
Fig. \ref{fig1}. (Fig. \ref{fig2}, respectively). 
Finally we note that we are primarily interested in non-radiative modes ($q>\omega/c$) in 
which case fields are localized near the graphene plane ($z=0$) and decay 
exponentially: $E(z),B(z) \propto e^{-|z|\sqrt{q^2-\omega^2/c^2}}$.

In order to find the plasmon-phonon coupled excitations we consider the 
complete Hamiltonian $H=H_e+H_{e-em}+H_{e-ph}+H_{ph}$. We assume that the 
hybrid plasmon phonon mode oscillates at some frequency $\omega$ 
with wavevector $q$ (which are to be found). From the equation of motion for 
the phonon amplitudes $Q_{\mu {\bf q}}$ one finds \cite{PinesBook}
\begin{equation}
(\omega^2 - \omega_0^2)\langle Q_{T {\bf q}} \rangle = 
L^2 F \langle { J}_{L}({\bf q},\omega) \rangle ,
\label{ph-resp-trans}
\end{equation} 
and 
\begin{equation}
(\omega^2 - \omega_0^2 ) \langle Q_{L {\bf q}} \rangle = 
-L^2 F \langle { J}_{T}({\bf q},\omega) \rangle .
\label{ph-resp-long}
\end{equation} 
The electron phonon interaction (\ref{e-ph2}) is included as 
the effective vector potential (\ref{Aeff}) in Eq. (\ref{e-em}), 
which from Eq. (\ref{current}) immediately yields 
\begin{equation}
\langle { J}_L({\bf q},\omega) \rangle = \chi_L({\bf q},\omega) 
( - \langle A_{L {\bf q}} (\omega) \rangle + F \langle Q_{T {\bf q}} \rangle),
\label{e-resp-long}
\end{equation}
and
\begin{equation}
\langle { J}_T({\bf q},\omega) \rangle = \chi_T({\bf q},\omega) 
( - \langle A_{T {\bf q}} (\omega) \rangle - F \langle Q_{L {\bf q}} \rangle).
\label{e-resp-trans}
\end{equation}
From Eqs. (\ref{ph-resp-trans}) - (\ref{e-resp-trans}) it is clear that transverse 
(longitudinal) phonons couple only to longitudinal (transverse) plasmons. 
Apparently, this follows from the fact that LO (TO, respectively) phonons are 
equivalent to oscillations of an effective vector potential ${\bf A}_{\textrm{eff}}$
[see Eq. (\ref{Aeff})], and therefore an effective electric field, perpendicular 
(parallel, respectively) to ${\bf q}$.

Finally using Eqs. (\ref{Maxwell-long}), (\ref{ph-resp-trans}), and 
(\ref{e-resp-long}) we get the dispersion relation for the 
LP-TO coupled mode
\begin{equation}
\omega^2 - \omega_0^2 = \frac{L^2F^2\chi_L({\bf q},\omega)}
{1-\frac{\sqrt{q^2-\omega^2/c^2}}{2\omega^2\epsilon_0}\chi_L({\bf q},\omega)},
\label{LP-TO}
\end{equation}
and from Eqs. (\ref{Maxwell-trans}), (\ref{ph-resp-long}), and (\ref{e-resp-trans}) 
dispersion relation for the TP-LO coupled mode
\begin{equation}
\omega^2 - \omega_0^2 = \frac{L^2F^2\chi_T({\bf q},\omega)}
{1+\frac{\mu_0}{2\sqrt{q^2-\omega^2/c^2}}\chi_T({\bf q},\omega)}.
\label{TP-LO}
\end{equation}
The plasmon dispersions relations (\ref{plasmon-long}) and (\ref{plasmon-trans}) appear 
as poles in the Eqs. (\ref{LP-TO}) and (\ref{TP-LO}) for the coupled modes, which means 
that the coupling is greatest at the resonance point where plasmon momentum and energy 
match that of the appropriate phonon mode. We denote this point (where the uncoupled plasmon 
and phonon dispersion cross) by $(q_c,\omega_0)$. One can quantify the strength of the 
coupling effect by calculating the frequency difference between the hybrid modes 
at the wavevector $q_c$ in units of the uncoupled frequency value: $\Delta\omega/\omega_0$. 
Finally by doping one can change plasmon dispersion which in turn changes $q_c$ and 
the strength of the plasmon-phonon coupling.

The dispersion lines for the hybrid LP-TO modes are shown in Fig. \ref{fig1} 
for two values of doping, (a) $n=5\times 10^{12}$~cm$^{-2}$, $E_F=0.261$~eV, $k_F=3.96\times 10^{8}$~m$^{-1}$, and 
(b) $n=5\times 10^{13}$~cm$^{-2}$, $E_F=0.825$~eV, $k_F=1.25\times 10^{9}$~m$^{-1}$. 
The strength of the coupling increases with increasing values of doping, and one has for the case
(a) $\Delta\omega/\omega_0=7.5 \% $, and (b) $\Delta\omega/\omega_0=15.5 \% $. 
To describe graphene sitting on a substrate (say SiC, which is a polar material), 
one only needs to include the dielectric function of the substrate into our calculation. 
In that case plasmons can also couple to surface phonon modes of the polar substrate \cite{Liu2010,Hwang2010,Koch2010}. 
However, since these surface phonons have sufficiently smaller energies than 
optical phonons in graphene out results are qualitatively unchanged in that case. 
LP-TO hybrid modes could be measured by observing the change in the phonon 
dispersion with the Neutron Spectroscopy or Inelastic X-ray Scattering.
Alternatively, one could use grating coupler or Electron Energy Loss 
Spectroscopy to measure the shift in the plasmon energy.
Our results imply that plasmon-phonon coupling could serve to explore 
the electron-phonon interaction (the frequency shifts at resonance are much 
larger then those recently measured by Raman Spectroscopy \cite{BO}), and that by 
externally appling voltage one can influence the properties of lattice vibrations.

In spite of the fact that the formal derivation of hybrid TP-LO coupled 
modes is equivalent to the derivation of the LP-TO modes, their 
properties qualitatively differ. First, we note that the dispersion of 
transverse plasmons is extremely close to the light line, and we 
plot $\Delta q=q-\omega/c$ vs. frequency $\omega$ following Ref. \cite{Mikhailov2007}. 
For this reason, transverse plasmons are expected to have strong polariton character 
and they will be hard to distinguish from free photons (also, even a small plasmon 
linewidth will obscure the distinction). 
Moreover, they do not exist in graphene between two dielectrics with sufficiently 
different relative permittivity, where the light lines for the dielectrics are 
separated.
Next, transverse plasmons exist only in the frequency interval 
$2E_F > \hbar\omega > 1.667 E_F$ \cite{Mikhailov2007}, which means that 
the LO phonon energy must be in the same interval  
for the hybridization to occur. 
Figure \ref{fig2} shows the dispersion curves of the hybrid TP-LO modes 
for two values of doping, 
(a) $n=7.5\times 10^{11}$~cm$^{-2}$, $E_F=0.101$~eV, $k_F=1.53\times 10^{8}$~m$^{-1}$, and 
(b) $n=9.5\times 10^{11}$~cm$^{-2}$, $E_F=0.114$~eV, $k_F=1.73\times 10^{8}$~m$^{-1}$. 
We observe that the trend here is opposite to that of the LP-TO coupling, as the strength of the coupling 
decreases with increasing doping; specifically, one has for the case 
(a) $\Delta\omega/\omega_0=0.17 \% $, and (b) $\Delta\omega/\omega_0=0.02 \%$.
The maximal coupling occurs when $2E_F$ is just above $\hbar \omega_0$, 
and it is zero when $\hbar \omega_0=1.667E_F$. 
We emphasize that the strength of the coupling for TP-LO modes is in general 
much weaker than in LP-TO modes.

Before closing, we note another interesting result which is captured by our calculations.
Equations (\ref{LP-TO}) and (\ref{TP-LO}) for shifts in 
the energies of TO and LO modes at $q=0$ reduce to 
\begin{equation}
\omega^2 - \omega_0^2 = \frac{L^2F^2\chi_{L,T}(0,\omega)}
{1+\frac{i}{2\omega\epsilon_0 c}\chi_{L,T}(0,\omega)},
\label{q0}
\end{equation}
which is identical to the result of Ref. \cite{Ando_anomaly}, where the 
coupling of optical phonons to single-particle excitations was studied, 
appart from the imaginary term in the denominator which is zero in \cite{Ando_anomaly}. 
This small but qualitative difference is consequence of phonon coupling to the 
radiative electromagnetic modes, which increases the phonon linewidth. 
For example, for the doping values of $n=5\times 10^{12}$~cm$^{-2}$, 
$5\times 10^{13}$~cm$^{-2}$, and $5\times 10^{14}$~cm$^{-2}$, Eq. (\ref{q0}) yields 
$0.005\%$, $0.07\%$, and $0.7\%$, respectively, for the 
linewidths, while there is no linewidth from single-particle damping 
at these doping values. This effect is qualitatively unchanged for graphene 
sitting on a substrate and could be measured by Raman spectroscopy. 
Finally, we note an interesting solution of Eq. (\ref{TP-LO}) (valid for suspended graphene): 
when the hybrid TP-LO mode dispersion crosses the light line it has the same 
energy as the uncoupled phonon mode, i.e.,  $\omega=\omega_0$. In other words, LO phonon at 
a wavevector $q=\omega_0/c$ decouples from all (single particle and collective) electron 
excitations, while no such effect exists for the TO phonons.

In conclusion, we have predicted hybridization of plasmons and intrinsic optical 
phonons in graphene using self-consistent linear response theory. 
To the best of our knowledge, this is the first study of such resonance 
in an isolated 2D material. 
We found that graphene's unique electron-phonon interaction leads to unconventional 
mixing of plasmon and optical phonon polarizations: 
longitudinal plasmons couple exclusively to transverse optical 
phonons, whereas graphene's transverse plasmons couple to longitudinal optical phonons;
this contrasts plasmon-phonon coupling in all previously studied systems. 
The strength of the hybridization increases with doping 
in LP-TO coupled modes, while the trend is opposite for TP-LO modes. 
The LP-TO coupling is much stronger than TP-LO coupling, 
and it could be measured by current experiments, which would 
act as a magnifier for exploring the electron-phonon interaction in graphene. 
This coupling is an even more striking example of a breakdown of Born-Oppenheimer 
approximation in graphene than the recently measured stiffening of the Raman G peak \cite{BO}. 
Moreover, plasmon-phonon interaction can serve to electronically control 
the frequencies of lattice vibrations in graphene, which could have 
interesting technological implications. In this context we should mention that our 
study opens the way for investigations of plasmon-phonon interaction 
in bilayer graphene, where phonon lasers were recently proposed \cite{Tang2010}.

This work was supported in part by the the Croatian Ministry of Science 
(Grant No. 119-0000000-1015), the MRSEC program of National 
Science Foundation of the USA under Award No. DMR-0819762. M.S. was also 
supported in part by the S3TEC, an Energy Frontier Research Center funded 
by the U.S. Department of Energy, Office of Science, Office of Basic 
Energy Sciences under Award No. DE-SC0001299.

%%%%%%%%%%%%%%%%%%%%%%%%%%%%%%%%%%%%%%%%%%%%%%%%%%%%%%%%%%%%%%%%%%%%%%%%%%%%%%%

\end{document}